# Observation of supermodulation in $LaO_{0.5}F_{0.5}BiSe_2$ by scanning tunneling microscopy and spectroscopy

Satoshi Demura, Naoki Ishida, Yuita Fujisawa and Hideaki Sakata

Tokyo University of Science, Department of Physics, Shinjyuku-ku, Tokyo 162-8601, Japan

E-mail : demura@rs.tus.ac.jp

**Abstract**

We observed surface and electronic structure of $LaO_{0.5}F_{0.5}BiSe_2$ single crystal by scanning tunneling microscopy/spectroscopy (STM/STS) at 4.2 K. Square lattice composed of Bi atoms was observed at a positive sample bias voltage on the surface prepared by cleavage. At a negative sample bias voltage, a stripe structure running along Bi-Bi directions was observed as in the previous report on $NdO_{0.7}F_{0.3}BiS_2$. Furthermore, we observed a supermodulation running along the diagonal directions with the period of about 5 times of the lattice constant. This seems to be indicative of structural instability of this system rather than electronic instability attributed to a nesting picture.

Recently, various kinds of layered BiS$_2$-based superconductors, $Ln$OBiS$_2$ (Ln = La, Pr, Ce, Nd, Yb, and Bi), have been discovered [1-27]. These materials have been intensively investigated because their layered structure, which is composed of superconducting layers and blocking layers, is similar to those of cuprates and iron-based superconductors. Although, the superconducting transition temperature ($T_c$) of these BiS$_2$-based materials is 3-5 K, $T_c$ strongly depends on external pressure or high-pressure annealing [2, 28-33]. For instance, LaO$_{1-x}$F$_x$BiS$_2$ showing $T_c$ of around 3 K shows superconductivity at around 10 K when an external pressure is applied. This higher $T_c$ remains at ambient pressure after as-grown polycrystalline samples are annealed under high pressure. These high-pressure effects indicate that the crystal structure of these materials is rather unstable against the external perturbations.

Theoretical calculations also predicted electronic and structural instability against the formation of charge density wave (CDW) [34-38]. A calculation of band structures predicted due to nesting at the F concentration of $x$=0.5, where the Fermi surface connects at around ($\pi$/2, $\pi$/2) in the $k$ space [34, 35]. The CDW is expected to be formed along the ($\pi$, $\pi$) direction in the $k$ space, whose direction corresponds to the diagonal directions of Bi square lattice in the real space with the period of about $\sqrt{2}$ times of the lattice constant. A calculated phonon dispersion curve found the existence of plateau at the negative energy between $\Gamma$ and M point in $k$ space, indicating a structural instability at F concentration of $x$=0.5 [35-38]. Although the direction of the instability agrees with that of the electronic instability, because of the plateau at the negative energy, the period of the CDW cannot be defined from the dispersion curve. These theoretical calculations suggest the superconductivity in BiS$_2$ based materials realizes near CDW ground state. Although the existence of the CDW was suggested by transport experiments in EuFBiS$_2$, this has not been confirmed by direct observations [10]. Thus, the relation between the superconductivity and the CDW is still unclear.

To investigate the existence of supermodulations such as CDW, microscopic real space observation is favorable. In fact, scanning tunneling measurements in $NdO_{0.7}F_{0.3}BiS_2$ and $CeO_{0.3}F_{0.7}BiS_2$ have revealed the existence of novel electronic structure called "checkerboard stripe" and "back bone" structure [39,40]. The "checkerboard stripe" structure consists of two dimensional alternating array of nano domains along the diagonal directions of Bi square lattice with about 5 time of a lattice constant. One nano domain consists of unidirectional electronic stripes along a Bi-Bi direction, whereas in the other nano domain stripes are running along the other Bi-Bi direction. These structures were observed at negative sample bias voltage, though tunneling spectroscopy measurements for these samples have revealed the existence of the large energy gap at Fermi energy ($E_F$), which is inconsistent with the metallic bulk properties such as temperature dependence of the electronic resistivity [41-43].

Very recently, modification of the superconducting layer by changing S atoms to Se atoms has been done [18]. The La(O,F)BiSe$_2$ has the same crystal structure as La(O,F)BiS$_2$ and shows superconductivity about 3 K. In this material, the same kind of CDW instability as in La(O,F)BiS$_2$ has been also predicted by a theoretical calculation [44]. In contrast, ARPES measurements have revealed more homogeneous distribution of the electronic states without any electronic inhomogeneity [45] than that of Nd(O,F)BiS$_2$. Thus, this material suits surface sensitive measurements to see the supermodulation. In this paper, we performed STM/STS measurements on La(O,F)BiSe$_2$ at 4.2K to investigate the surface and the electronic state, and compared the result with those in Nd(O,F)BiS$_2$. We observed square lattice of Bi atom on the surface prepared by cleavage. Tunneling spectroscopy measurements revealed the existence of a finite local density of states at Fermi energy in $LaO_{0.5}F_{0.5}BiSe_2$, which is consistent with the bulk properties such as temperature dependence of resistivity. This finding opens the possibility to examine the superconducting gap or electronic states in vortices to determine the symmetry of the superconductivity and lead the elucidation of the superconducting mechanism in this material by STM/STS. At a negative sample bias voltage, stripe structure was observed as in the previous report on $NdO_{0.7}F_{0.3}BiS_2$. Furthermore, an electronic supermodulation with the period of about 5 times of the lattice constant was observed at the negative sample bias voltage along the diagonal directions of Bi square, whose directions correspond to the theoretically predicted directions of CDW. However, the period of the observed supermoduration is inconsistent with the predicted CDW. Therefore, this seems not to be described by the nesting picture.

Single crystal samples of $LaO_{0.5}F_{0.5}BiSe_2$ were synthesized by a CsCl flux method in vacuumed quarts tubes [46,47]. Mixtures of Bi (Mitsuwa Chemicals Co. Ltd., 99.9%), Se (Kojyundo Chemical Laboratory Co. Ltd., 99.99%), $Bi_2O_3$ (Kojyundo Chemical Laboratory Co. Ltd., 99.99%), and $BiF_3$ (Stella Chemifa Co. Ltd., 99%) were ground with nominal compositions of $LaO_{0.5}F_{0.5}BiSe_2$ except for La. After grind, La grains (Rare Metallic Co. Ltd., 99.9%) were put into this mixture. The mixture of 0.8 g was mixed with CsCl powder (Kojyundo Chemical Laboratory Co. Ltd., 99.9%) of 5 g which was annealed at 200 °C for 12 hours before the mixing. These mixture were sealed in an evacuated quartz tube. The tube was heated at 800 °C for 10 hours and cooled down to 630 °C at a rate of 0.5 °C/h. After this thermal process, the flux was removed by washing with distilled water. The obtained single crystals were confirmed to be $CeOBiS_2$ type structure with the space group *P*4/*nmm* symmetry by X-ray diffraction measurements with Cu-K$\alpha$ radiation using the $\theta$-$2\theta$ method. The surface structure of single crystals was observed by a laboratory-build scanning tunneling microscope (STM) in the He gas at 4.2 K. The clean surface was prepared by cleaving the single crystal at 4.2 K in situ. A bias voltage was applied to the sample in all measurements. d*I*/d*V* spectra were obtained by numerically differentiation of *I-V* characteristics.

Figure 1(a) shows a typical STM image on a cleavage surface taken at a positive sampel bias voltage. Because the crystal structure of this material has a Van der Waals gap between two $BiSe_2$ layers, the observed surface is one of the $BiSe_2$ plane. This surface shows a square lattice with a period of about 0.5 nm ($a_0$) as shown in the inset of Figs. 1(a). The observed atoms are thought to be Bi atoms as in the case of $NdO_{0.7}F_{0.3}BiS_2$ and $CeO_{1-x}F_xBiS_2$ ($x$=0.5, 0.7) [39,40]. Bi defects observed as black spots and streaks running along the diagonal directions are also observed in this material, though the number of defects is smaller than that of any other $BiS_2$-based materials [48].

Figure 1(b) shows a STM image taken at a negative sample bias voltage at the same filed view of Figs. 1(a). In this image, in addition to Bi atoms, one dimensional stripe structures running along Bi-Bi directions which are not observed at the positive sample bias voltage can be seen. Figure 1(c) shows a magnified image of the area shown in the white square in Figs. 1(b). As can be seen in this figure, there are nano domains. In one nano domain, the stripes at a length of approximately 4~5 times of the lattice constant $a_0$ are running along one of Bi-Bi directions, whereas in the other cluster the stripes are running along the other Bi-Bi direction. The average size of these clusters is about 2 $nm^2$. Such nano domains were also observed in $BiS_2$-based materials [39,40]. Thus, the observed stripe structure is inherent not only in $BiS_2$-based materials but also in $BiSe_2$-based materials.

In $BiS_2$-based materials, the nano domains formed "checkerboard structure", which is the square lattice composed of two alternating nano domains in which the stripes are running along the one of the Bi-Bi directions. In contrast, the nano domains in $LaO_{0.5}F_{0.5}BiSe_2$, does not show such checker board order, but shows rather disordered configuration. Instead, in addition to the stripes, a supermodulation superimposed on the stripes was observed only at the negative sample bias voltage. The low pass filtered image of Figs. 1(b) which elucidates the supermodulation is shown in Figs. 1(d). Although the supermodulation is not regular, the period is seems to be about five times of the lattice constant along the diagonal directions of Bi square lattice, which is discussed later. The supermodulation seems not to be two dimentional but unidirectional with changing the direction locally from one of the diagonal directions of Bi square to another with rather short coherence length, breaking four-fold symmetry. Short coherence length may indicate that the observed supermodulation is not necessarily long range order but represents structural instability of this system against a supermodulation with this wave vector. This is consistent with the lack of clear anomaly in transport measurements which indicate the existence of a phase transition [17,46].

In order to confirm the period of the modulation, we show the Fourier transformed image of Figs. 1(b). Dark spots pointed by Orange and Blue arrows in Figs. 2(a) correspond to Bragg spots of the Bi square lattice. There are satellite spots near the Bragg spots. These spots are considered to reflect the supermodulation along each direction. The average $\delta q$ in both directions is ~$0.19 \times 2\pi/a_0$ in the $k$ space, indicating the period is about 5 $a_0$ for both directions in real space.

Next, a spatial change in spectrum was investigated. Figure 2(d) shows a spatially averaged d$I$/d$V$ spectrum (solid line) on the surface of $LaO_{0.5}F_{0.5}BiSe_2$ single crystal. The asymmetric spectrum with a high conductance at the positive sample bias voltage is observed. This asymmetric shape of the spectrum is same as that of previous results in $NdO_{0.7}F_{0.3}BiS_2$ [39]. Thus, this shape is common feature in both $BiS_2$ and $BiSe_2$-based materials. On the other hand, $LaO_{0.5}F_{0.5}BiSe_2$ shows finite LDOS at $E_F$, whereas $NdO_{0.7}F_{0.3}BiS_2$ does not. Finite LDOS at $E_F$ is consistent with the metallic resistivity in $LaO_{0.5}F_{0.5}BiSe_2$, indicating the STM/STS measurements safely represent bulk properties of this sample. Dashed and dotted curves in Figs.2 (d) indicate the averaged spectra at brighter and darker regions in the Figs. 1(d), respectively. As can be seen, the change in d$I$/d$V$ between brighter and darker region is rather uniform, indicating the LDOS below $E_F$ uniformly contributes to the supermodulation.

Let us consider a relation between the observed supermodulation and the theoretically predicted CDW. In some theoretical calculations, possible CDW that is caused by Peierls mechanism has been proposed. From these theoretical calculations in LaO$_{1-x}$F$_x$BiSe$_2$, following features are expected.

(i) The instability against CDW formation is expected to develop along ($\pi,\pi$) direction in the $k$ space [44].

(ii) This instability of CDW is enhanced at the F concentration of 0.5, where Fermi surfaces connect [34-37].

(iii) A period of the predicted CDW is expected to be $\sqrt{2}$ times of the lattice constant $a_0$, which is corresponding to the nesting vector along ($\pi,\pi$) direction.

The observed supermodulation is consistent with features (i) and (ii). On the other hand, the observed period of about 5 times of the lattice constant $a_0$ is inconsistent with (iii). Furthermore, if CDW caused by Peierls mechanism were realized, clear phase transition to the CDW phase and formation of two dimensional supermodulation would be expected. However, both features were not observed although anomalous hamp-behavior in the temperature dependence of resistivity has been observed [46, 49, 50]. These results indicate that the observed supermodulation is not simple CDW due to Peierls mechanism.

Apart from the nesting picture, the calculation of phonon dispersion curve in Bi$Ch_2$ materials shows plateau at negative energy between Γ and M point in $k$ space, indicating lattice instability at any wave vector along this direction. Thus, the period of the observed supermodulation is possibly to deviate from $\sqrt{2}$ times of the lattice constant by the lattice instability. This suggests that there are some factors to stabilize the supermodulation period of 5 $a_0$ in Bi$Ch_2$ materials. Recently, a supermodulation with the same period of 5 $a_0$ has been reported in the synchrotron X-ray diffraction measurements in LaO$_{1-x}$F$_x$BiS$_2$ [51]. This result also supports the stability of the supermodulation with the period of 5 $a_0$ in these materials.

It is also noted that the period of the observed supermodulation is close to that of "checkerboard structure" observed in NdO$_{0.7}$F$_{0.3}$BiS$_2$. In addition, the "checkerboard structure" was only observed at the negative sample bias voltage. A period of this structure is around 5 $a_0$. These suggest that "checkerboard structure" observed in NdO$_{0.7}$F$_{0.3}$BiS$_2$ has the same origin as the supermodulation in LaO$_{1-x}$F$_x$BiSe$_2$.

In summary, we successfully observed the surface and electronic structure of single crystals for LaO$_{0.5}$F$_{0.5}$BiSe$_2$. The stripe structures aligned along two Bi-Bi directions was observed in the image of a negative sample bias voltage, indicating the stripe structure is inherent in BiSe$_2$ layer as well as BiS$_2$ layer. In contrast to BiS$_2$-based materials, finite LDOS at $E_F$ was observed. This finding opens for measurements of superconducting properties such as superconducting gap or vortices to elucidate the superconducting mechanism. Furthermore, a supermodulation along the diagonal directions of Bi square was observed at a negative sample bias voltage. Since the observed period of the supermodulation does not correspond to that expected from the nesting vector, the observed supermodulation is not simple CDW due to Peierls instability, but due to the lattice instability.

**Acknowledgements**

This work was partly supported by a Grant-in-Aid for Young Scientists (B) (No. 15K17710) in a Grant-in-Aid

**Figure captions**

Figure 1 (a), (b) STM images at a positive and negative sample bias voltage at the same field of view taken at $V_{set}$ = +500 and -500 mV on the cleaved surface of $LaO_{0.5}F_{0.5}BiSe_2$, respectively. The set tunneling current $I_{set}$ of both images is 300 pA. Inset in (a) shows the image at the smaller field of view than that of Figs. 1(a). (c) Magnified image in the white square shown in Figs. 1(b). Stripes along two directions $a_x$ and $a_y$ are shown as orange and blue square, respectively. (d) Low pass filtered image of Figs.1(b). The cut-off frequency is $|q| = 0.7 \times 2\pi/a_0$, which corresponds to the real-space cut-off radius of about 1.4 $a_0$.

Figure 2 (a) Fourier transformed (FT) image of Figs. 1(b). FT peaks corresponding to stripes along two directions $a_x$ and $a_y$ are shown as orange and blue circle, respectively. (b,c) Magnified figures of Figs. 2(a) near $q_y$ and $q_x$ spots, respectively. $\delta q$ shown by the arrow indicates the wave vector of the supermodulation. (d) The spatially averaged conductance spectrum (solid line) taken at $V_{set}$ = 600mV and $I_{set}$ = 400 pA on the surface of $LaO_{0.5}F_{0.5}BiSe_2$ single crystal. The dotted and dashed lines exhibit a spatially averaged conductance in the brighter and darker region in the Figs. 1(d), respectively.

Figure 1

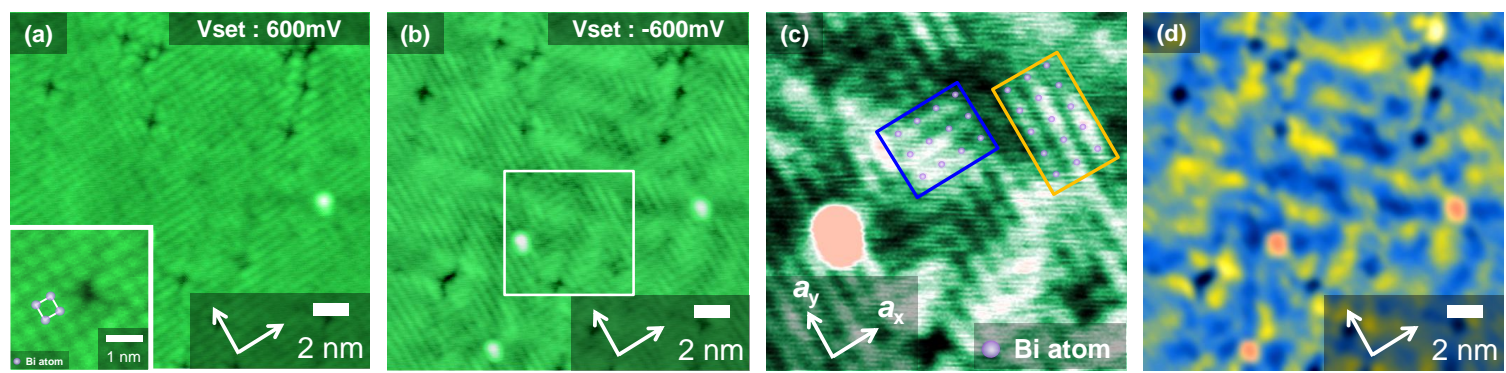

Figure 2

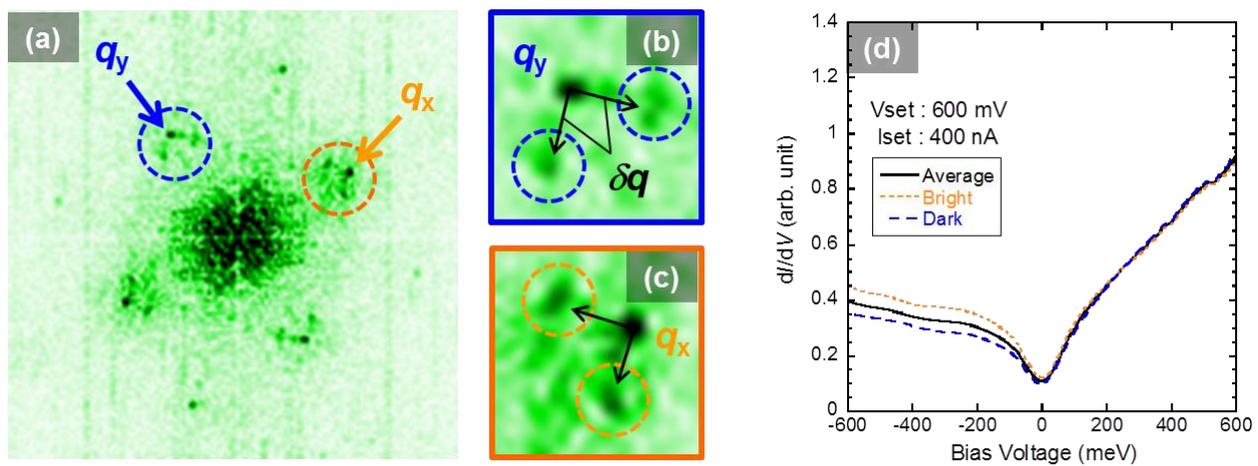